\documentclass[letterpaper,11pt]{article}

\usepackage{graphicx}
\usepackage{amssymb}
\usepackage{amsmath}
\usepackage[usenames,dvipsnames]{color}
\usepackage{hyperref}
\usepackage[utf8x]{inputenc}
\usepackage[left=1in,right=1in,top=1in,bottom=1in]{geometry}   
\hypersetup{
    colorlinks=true,         
    linkcolor=blue,          
    citecolor=red,        
    urlcolor=Violet             
}

\title{Quantisation, Representation and Reduction; How Should We Interpret the Quantum Hamiltonian Constraints of Canonical Gravity?}

\author{Karim P. Y. Th\'ebault\footnote{Centre for Time, School of Philosophical and Historical Enquiry, University of Sydney, Australia; E-mail: karim.thebault@gmail.com.}}

\begin{document}
\maketitle

\abstract{Hamiltonian constraints feature in the canonical formulation of general relativity. Unlike typical constraints they cannot be associated with a reduction procedure leading to a non-trivial reduced phase space and this means the physical interpretation of their quantum analogues is ambiguous. In particular, can we assume that `quantisation commutes with reduction' and treat the promotion of these constraints to operators annihilating the wave function, according to a Dirac type procedure, as leading to a Hilbert space equivalent to that reached by quantisation of the problematic reduced space? If not, how should we interpret Hamiltonian constraints quantum mechanically? And on what basis do we assert that quantisation and reduction commute anyway? These questions will be refined and explored in the context of modern approaches to the quantisation of canonical general relativity.}


\newpage
\tableofcontents

\section{Introduction}

Since their development in the late nineteen fifties the mathematical foundations of both the constrained Hamiltonian theory of mechanics and the constraint quantisation programme \cite{Dirac1964,HenTeit1992} have been substantially clarified. In particular, geometric insights into both mechanics \cite{Butt2007,GotNesHin1978,Sour1997} and quantisation \cite{EEML1999,Lands2007} have afforded a degree of precision and rigour in the canonical characterisation of gauge systems at both classical and quantum levels which was hitherto unavailable. An important formal result that has been established is that for a certain classes of constrained systems it can be proved \cite{GS1982,Gotay1986,BGMW1995,ConFre2009} that the Dirac route of first promoting the classical constraints to quantum operators \textit{and then} imposing them as restrictions on the physical wavefunctions should be considered equivalent to \textit{first} reducing out their action and then quantising the reduced phase space which results. On a conceptual level such proofs that `quantisation commutes with reduction' secure a \textit{representative equivalence} between the Hilbert space constructed via the Dirac method and that associated with the reduced phase space in the sense that they guarantee that: i) the action of the same unphysical gauge group has been mitigated for; ii) an identical reduced set of \textit{physical} degrees of freedom have been realised quantum mechanically. Furthermore they also establish that the two observable operator algebras of can thought of as equivalent in terms of representing the same ontology\footnote{If such a `commutation' proof is available then the observables will be connected by a far stronger relationship of being unitarily equivalent -- however for our purposes a weaker representational notion will prove more useful}.     

Dirac's major purpose in constructing both constrained Hamiltonian theory and constraint quantisation was to find a methodology for quantising gravity. However, on both a technical and conceptual level application of both aspects of his scheme have been found to be problematic. To a large degree these problems derive their origin from the complex and perplexing role of the Hamiltonian constraints within the canonical version of general relativity. These long standing issues are collectively known as the problem of time in quantum gravity and were reviewed in detail almost twenty years ago \cite{Kuchar1991,Isham1992}. Since then some significant advances have been made \cite{Ander2010}, but on a conceptual level  we still want a satisfactory resolution to the most thorny problem of all -- how should we interpret the action of the Hamiltonian constraints on a quantum level. To this author's knowledge, all existent approaches follow the Dirac convention in treating quantum Hamiltonian constraints (or something equivalent to them) as operators annihilating the wavefunction. However, if we are to assume that `quantisation commutes with reduction', then this is equivalent to endorsing the reduced phase space of the \textit{classical} theory as the fundamental dynamical arena. This is highly problematic since investing in such a space would seem to do violence to the physical aspect of the transformations generated by the Hamiltonian constraint and therefore render our dynamics trivial. The Hamiltonian constraints of canonical general relativity undoubtably have a different geometric origin to that of standard first class constraints.\footnote{In particular their phase space action cannot be unambiguously connected to the cotangent bundle projection of a symmetry group acting on the tangent bundle see \cite{PonSalSun2010} and references therein.} What is underappreciated is that this, together with the manifestly dynamical aspect to their phase space action, indicates that the standard interpretational framework for first class constraints should be no longer be thought automatically applicable. In fact, contra Dirac, these constraints, although first class, cannot be taken as generating \textit{phase space} transformations which `do not change the physical state'\footnote{This crucial point has been consistently made by both Kucha\v{r} \cite{Kuch1992} and Barbour \cite{Bar1994}. The extent to which the action of the constraints can be interpreted as gauge within a different context will be discussed in \S3.2} and therefore the quotienting out of their action will lead to a reduced space without a legitimate physical basis. Any approach to quantum gravity that makes use of a Hilbert space equivalent to that based upon the quantisation of this reduced space is therefore built upon dubious physical foundations.        

The complex structure of both the Hamiltonian constraints and the associated constraint algebra is such that they lie outside the scope of existent commutation proofs. Thus, on a formal level we are not justified in questioning current constraint quantisation techniques due to formal equivalence with a trivialising reduction. However, on a conceptual level we may be able to establish that the Hilbert space constructed by application of the modern variants of the Dirac constraint quantisation approach is representatively equivalent to that based upon quantisation of the problematic reduced space. Such an argument, if it can be constructed, would give us cause to question the interpretative basis of the Wheeler-De Witt equation and, moreover, any theory of quantum gravity that follows the Dirac line on quantisation, in particular loop quantum gravity.

\section{Reduction and quantization}
\subsection{Symplectic reduction and geometric quantization}
In phase space terms, the existence of a gauge symmetry group corresponds to the
$p_{i}$'s and $q_{i}$'s not all being independent -- there exists some
functional relationship between them of the form $\varphi (p,q)=0$. We call
such functions \textit{constraints}\footnote{These particular constraints are \textit{primary} constraints since they arise directly from the definition of the canonical momenta. It is also possible that there exist secondary constraints but for the purposes of this paper this distinction will be unimportant} and when a given constraint commutes with all the others we call it \textit{first class}. For our purposes it will be safe to assume that all constraints are first class. Geometrically, we can understand the collection of all the (first class) constraints, $%
\varphi _{j}$ $j=1,...m$, as defining an $m$ dimensional sub-manifold, $%
\Sigma =\{(p,q)\in \Gamma|\forall _{j}:\varphi _{j}(p,q)=0\},$ within the extended 
phase space, $\Gamma $, that we call the constraint surface. The extended phase
space itself has a symplectic geometry
characterised by the pair $(\Gamma ,\Omega )$ -- where $\Omega $ is
a closed and non-degenerate two form which we assume, for simplicities sake, to be constructed by taking the total
differential of the Poincar\'{e} one-form $\theta =p_{i}dq^{i}$. Points in this space which do not lie on the constraint surface will not correspond to physically possible states since they constitute solutions
which violate the gauge symmetry. We therefore designate the constraint surface the \textit{physical phase space}.  The geometry of the physical phase space is given by
first restricting $\theta $ to $\Sigma $ to get a new characteristic one-form, $\tilde{\theta}=\theta _{|_{\Sigma }}$. The total derivative of $%
\tilde{\theta}$ will then give us a two-form $\omega=\mathbf{d}%
\tilde{\theta}$, which endows the constraint manifold with the geometry $%
(\Sigma ,\omega)$. This new two form will be closed but whether it
is degenerate or not depends on the particular properties of the physical phase space itself. For the typical case  where $\omega$ is degenerate our physical phase space will have a \textit{%
presymplectic} geometry. 

Presymplectic geometries have a degenerate structure that does not allow us to associate a unique vector field with every smooth function\footnote{Symplectic geometries on the other hand equip us with a Poisson bracket structure such that $\Omega(X_{f},\cdot)=\{f,\cdot\}=\mathbf{d}f$ gives a unique $X_{f}$ for every $f\in C^{\infty}(\Gamma)$}. This means the conventional methodology for constructing dynamics in terms of the integral curves of the vector field associated with the Hamiltonian function will not lead to unique results. A form of indeterminism is present such that multiple mathematically distinct yet dynamically equivalent solutions are defined at a given point irrespective of the path leading up to that point. To relieve ourselves of the interpretation difficulties that this breakdown of the correspondence between solutions and physical histories creates we can remove the degeneracy within our physical phase space by following a symplectic reduction procedure leading to a reduce phase space.  

Let us define the null tangent vector space $N_{x}\subset T_{x}\Sigma $ as the collection of vectors that satisfy the equation $\omega(X,\cdot )=0$. This is equivalent to the null space or kernel, $Ker(\omega)$, of the
presymplectic form. An equivalence relation between two points $x,y\in \Sigma $ can
be defined based upon the condition of being joined by a curve, $\bar{%
\gamma}:%
\mathbb{R}
\rightarrow \Sigma $, with null tangent vectors. Sets of points for which
this equivalence relation holds are sub-manifolds called gauge orbits, $[x]$,
and we say that the action of our presymplectic form is to partition phase
space into these orbits. Equivalently, the orbits are defined
by the integral curves of the null vector field of $\omega$. The non-uniqueness that we understood in terms of the
existence of gauge orbits is, therefore, also characterised by $Ker(\omega)$. Critically for our purposes, the quotient $\Pi _{R}=\Sigma /Ker(\omega)$ will necessarily be, i) a manifold and, ii) have symplectic structure.\footnote{The first is assured since the quotient we are taking is of a manifold by a \textit{sectional foliation} see \cite{Sour1997} p.42 and pp.82-83. It is a sectional foliation because the orbits which partition $\Sigma$ constitute manifolds which are suitably transverse. The second is assured because the quotient is of a presymplectic manifold with respect to the kernel of its own presymplectic form and it can be shown that this implies that the resulting quotient manifold will be endowed with a closed two form with a kernel of zero dimension -- i.e. it will have a symplectic geometry see  \cite{Sour1997} theorem 9.10.} We can now represent dynamics uniquely within the the reduced phase space $\Pi_{R}$  since the projection
map $\pi :\Sigma\rightarrow \Pi _{R}$ defines a Hamiltonian system $%
(\Pi _{R},\omega _{R},H_{R})$ where $H_{R}$ is taken to be a non-trivial Hamiltonian function on the reduced space and $\omega _{R}$ is the two form whose
pullback to $\Sigma $ by $\pi $ is $\omega$ (i.e. $\omega=\pi^{\ast}\Omega _{R}$ where $\pi^{\ast }:\Pi _{R}\rightarrow\Sigma$). An equation of the
form $\Omega _{R}(X_{H_{R}},\cdot )=\mathbf{d}H_{R}$ then gives us a unique
Hamiltonian vector field with which to define dynamical solutions in one-to-one correspondence with physical histories. This unique representational capacity allows us to treat functions on the reduced phase space, $\mathcal{O}_R\in C^{\infty}(\Pi_{R})$, as representing physical observables -- since  such functions will be both uniquely and deterministically defined on any dynamical solution. We can also construct an analogous restriction on functions on the original extended phase space, $\Gamma $, by use of the fact that the constraints generate vector fields which are null on the constraint surface. This means that any function which weakly commutes (i.e. has zero Poisson bracket on the constraint surface) with the constraint functions must be constant along gauge orbits and therefore that such a function is uniquely and deterministically defined on dynamical solutions. Thus, we have $\{\mathcal{O},\varphi _{j}\}|_{\Sigma}=0$ for $\mathcal{O}\in C^{\infty}(\Gamma)$       

The objective of the geometric quantisation programme \cite{EEML1999} is to find a correspondence between the sets of pairs constituted by: symplectic manifolds $(\mathcal{M},\Omega)$  together with smooth real functions $C^{\infty}(\mathcal{M})$, on the one hand; and complex Hilbert spaces $\mathcal{H}$ together with self-adjoint operators $\mathcal{A}(\mathcal{H})$, on the other. 
We define the full quantisation of a classical system $(\mathcal{M},\Omega)$ as a pair $(\mathcal{H}_{Q},A)$ under certain conditions on $\mathcal{H}_{Q}$ and the map, $A$, which takes us between classical and quantum observables.\footnote{Explicitly we require that: 1) $\mathcal{H}_{Q}$ is a separable complex Hilbert space. The elements $\mid \psi \rangle \in \mathcal{H}_{Q}$ are the quantum wavefunctions and the elements $\mid \psi \rangle_{\mathbb{C}}\in \mathbf{P} \mathcal{H}_{Q}$ are the quantum states where $\mathbf{P} \mathcal{H}_{Q}$ is the projective Hilbert space;
2) $A$ is a one to one map taking the classical obervables $f\in \Omega ^{0}( \mathcal{M})$ to the self adjoint operators $A_{f}$ on $\mathcal{H}_{Q}$ such that:  i) $A_{f+g}=A_{f}+A_{g}$ ii) $A_{\lambda f}=\lambda A_{f}$ $\forall \lambda \in \mathbb{C}$ iii) $A_{1} = Id_{\mathcal{H}_{Q}}$; 
3) $[A_{f},A_{g}]=i\hbar A_{\{f,g\}}$ (i.e. $A$ is a Lie algebra morphism up to a factor);     
4) For a complete set of classical observables $\{f_{j}\}$, $\mathcal{H}_{Q}$ is irreducible under the action of the set $\{A_{f_{j}}\}$.} We can see this quantisation programme as consisting essentially of the construction of a Hilbert space $\mathcal{H}_{Q}$ on which the Lie algebra of classical observables can be represented irreducibly in terms of a set of self-adjoint operators $\mathcal{A}(\mathcal{H}_{Q})$ -- the elements of this set are the quantum observables. When combined with the symplectic reduction procedure outlined above, geometric quantization gives us a methodology for quantising a system with first class constraints -- i.e. first reduce then geometrically quantise the reduced phase space making use of the symplectic structure that our reduction procedure guarantees. Explicitly, what we do is consider the reduced phase space with geometry $(\Pi_{R},\Omega_{R})$ and set of reduced observable functions  $\mathcal{O}_R$ to be our classical pairing and find the corresponding Hilbert space $\mathcal{H_{R}}$ and self-adjoint operators $\mathcal{\hat{O}_{R}}(\mathcal{H_{R}})$. If the symplectic reduction procedure runs through successfully we are guaranteed to be able to construct the quantum equivalent.      

\subsection{Extended phase space quantization and quantum constraint reduction}

The key move in Dirac's constraint quantisation approach \cite{Dirac1964,HenTeit1992} is to quantise the extended phase space $\Gamma$ such that the physical phase space of a canonical gauge theory sits within. As we have seen this space will have a symplectic structure. We can therefore promote smooth phase space functions, $f\in C^{\infty}(\Gamma)$, into Hermitian operators, $\hat{f}$, and the Poisson bracket relation, $\{f,g\}=\Omega(X_{f},X_{g})$, into commutation relations with the appropriate $i\hbar$ factors. This essentially amounts to a partial application of the geometric procedure above. The Hilbert space that results is called the auxiliary Hilbert space $\mathcal{H}_{aux}$ and we can define a class of auxiliary state vectors $\mid \psi_{aux} \rangle$. We then impose the (first class) constraint functions as operators on $\mathcal{H}_{aux}$ restricting the physical state vectors $\hat{\phi}_{j}\mid \psi_{phys} \rangle=0$. The Hilbert space that is constructed by taking the physical states is the physical Hilbert space $\mathcal{H}_{phys}$ of the quantum theory. We are provided with a set of quantum observables $\hat{\mathcal{O}}$ by considering the set of self-adjoint operators which commute with the constraints and map physical states to physical states.

Formally the quantisation procedure we have just sketched suffers from a number of difficulties and ambiguities not least: 1) the quantisation of the classical constraint functions on phase space is not unique due to a factor ordering ambiguity; 2) extra input is needed to define a Hilbert space structure on the physical states in particular an inner product; 3) solving the constraints at the quantum level is non-trivial and may lead to inconsistent results.\footnote{There is also the additional problem that if the constraints depend non-polynomially on the field variables then it may prove impossible to find a rigorously defined representation of them on the $\mathcal{H}_{aux}$. This issue is particularly pressing for the constraints of canonical general relativity and leads, in that case, to the introduction of Ashtekar variables. However, neither this formal issue, nor the structure of the new variables, have any particular bearing on the our more conceptual concerns regarding the nature of quantum Hamiltonian constraints. Their discussion can, therefore, be reasonably neglected for the purposes of this non-explicit treatment.}   

A number of modern strategies are available that allow us to formalise the Dirac quantisation scheme such that these issues can be overcome or at least diminished. The two that we will briefly consider here are the \textit{group averaging methodology} (as used in refined algebraic quantisation) and the \textit{direct integral approach} (as applied in the master constraint programme). These two techniques are particularly significant for our purposes since a combination of them is utilised in the loop quantum gravity approach to quantising general relativity \cite{Thie2007}. 

Refined algebraic quantisation \cite{GiulMar1999a,GiulMar1999b} (RAQ) is a methodology for addressing the ambiguities of the Dirac quantisation scheme whilst still staying within the broad outline of `quantise first, constrain second'. As per the original Dirac approach, we first construct a Hilbert space representation of the operator algebra of functions on the extended phase space. The constraints are then taken to be represented as Hermitian operators acting on this $\mathcal{H}_{aux}$. Crucially, we require that the the commutator algebra of the quantum constraints forms a Lie algebra (this will always be the case provided the classical Poisson bracket constraint algebra closes with structure constants) -- exponentiation of the constraint operators will then yield a unitary representation $U(g)$ of the corresponding Lie group $G$. Let us then define some subspace $\Phi\subset \mathcal{H}_{aux}$ together with its algebraic dual $\Phi^{\star}$ (i.e. the space of complex valued linear functions $f$ on $\Phi$). If the space $\Phi$ is chosen such that the constraint operators map it into itself then a well defined dual action of these operators is also available.\footnote{i.e. we have that $U(g)f[\phi]=f(U{g^{-1}\phi})$, $\forall \phi\in\Phi$} Solutions of the constraints are then elements $f\in\Phi^{\star}$ for which $U(g)f=f$  $\forall g\in G$. Physical observables can then be defined as self adjoint operators $\mathcal{\hat{O}}(\mathcal{H}_{aux})$ which include $\Phi$ in their domain, map $\Phi$ to itself and (crucially) commute with the group action on $\Phi$.\footnote{$\mathcal{\hat{O}}U(g)\mid \phi \rangle=U(g)\mathcal{\hat{O}}\mid \phi \rangle$, $\forall g\in G$, $\phi\in\Phi$} 

 The pivotal move is the definition of the \textit{rigging map} which is an anti-linear map $\eta$ from $\Phi$ into $\Phi^{\star}$ such that: its image solves the constraints; it is real and positive; it commutes with the observables. The RAQ scheme then provides us with a methodology for constructing the physical Hilbert space since an inner product is provided to us by the rigging map:  $\langle \eta(\phi_{1}),\eta(\phi_{2})\rangle_{phys}=\eta(\phi_{1})[\phi_{2}]$. This new inner product is defined on $\Phi$ and it leads us to the physical Hilbert space $\mathcal{H}_{phys}$ via taking the quotient of $\Phi$ by the sub-set of zero norm vectors it defines. The physical observables will then be automatically defined as operators on $\mathcal{H}_{phys}$ and the correspondence between the RAQ definition of observables and the original Dirac one given above, becomes explicit.

	Clearly, the success of RAQ depends on our ability to find a suitably unique rigging map. This can be done subject to the restriction that $G$ is a locally compact Lie group with a Haar measure $\mu_{H}$.\footnote{A right (left) Haar measure is a positive measure on a group invariant under right (left) translations. For the uni-modular case which we are restricting ourselves to, the left and right Haar measures agree.} In these circumstances (and for the case that the group is unimodular see \cite{GiulMar1999b} for the non-unimodular case) then the the group averaging methodology defines the rigging map simply as: 
	\begin{equation}
\eta\mid\phi\rangle:=\langle\phi\mid\int d\mu_{H} U(g)
\end{equation}
That this rigging map solves the constraints is guaranteed by the invariance of the Haar measure  and that it is real and commutes with the observables is guaranteed by the fact that it is invariant under $g\rightarrow g^{-1}$. 	 

The Master Constraint Programme\cite{Thie2007,Thie2003,DitThe2006} (MCP) for the quantisation of constrained systems constitutes more of a departure from the Dirac scheme than RAQ since it leads us to a different representation of the constraint functions even at a classical level. It is still of the Dirac quantisation genus, however, since these reformulated constraints are again only imposed after quantisation. A particular strength of the the MCP approach is that it remains well defined even for systems where the Poisson bracket algebra of the constraints closes only with structure functions. This is particularly important feature for our purposes since the Hamiltonian constraints of canonical general relativity are associated with an algebra of exactly this type. 

The essential idea is to re-write the classical constraint functions, $\varphi _{j}(p,q)=0$, in terms of a single equation which will be satisfied under the same conditions. This new single constraint is then the \textit{Master Constraint}. A simple example is given by taking a positive quadratic two form $K^{ij}$ and constructing the equation:
\begin{equation}
\mathbf{M}:=K^{ij} \varphi _{i}\varphi _{j}=0
\end{equation}
This equation is satisfied if and only if all the individual constraint functions are vanishing and thus defines the same physical phase space $\Sigma$ that we had before. We can recover our observable condition for the \textit{extended phase space} by considering the class of functions such that:
\begin{equation} 
\{\{\mathbf{M},\mathcal{O}\},\mathcal{O}\}|_{\mathbf{M}=0}=0
 \end{equation}
i.e. those functions which have a vanishing double Poisson bracket with the master constraint on the constraint surface. The geometric interpretation of this condition on classical observables is subtly, yet importantly, different to the standard one given above. Strictly, it is a restriction that implies that the observable functions generate finite symplectomorphisms which preserve $\Sigma$, rather than the usual condition that the observables are constant along the null directions generated by the constraints on  $\Sigma$. However, it can be straight forwardly demonstrated that the two conditions are equivalent \cite{Thie2003}. Thus the intuitive connection between these observables and the $\mathcal{O}_{R}$ of \S2.1 is retained. We may therefore think about the $\mathcal{O}$ as corresponding to functions projected up from the reduced phase space. 
  
Moving on to quantisation, we look for a representation of the Poisson algebra of functions on the extended phase space, $f$, in terms of commutator algebra of operators, $\hat{f}$, on a (separable) auxiliary Hilbert space $\mathcal{H}_{aux}$. We then require that the Master Constraint $\mathbf{M}$ is represented as a positive, self-adjoint operator $\hat{\mathbf{M}}$. This is possible even if the classical constraints cannot themselves be represented in such a way -- i.e. when they fail to form a Lie algebra under the Poisson bracket operation. Following Thiemann \cite{Thie2007,Thie2003}, since $\mathcal{H}_{aux}$ is by assumption a separable Hilbert space it can be represented as a direct integral of separable Hilbert spaces $\mathcal{H}_{aux}^{\oplus}(\lambda), \lambda\in\mathbb{R},$ subordinate to $\hat{\mathbf{M}}$ according to:
\begin{equation}
\mathcal{H}_{aux}=\int_{\mathbb{R}}^{\oplus} d\nu(\lambda)\mathcal{H}_{aux}^{\oplus}(\lambda)
\end{equation}
where although the measure $\nu$ and Hilbert spaces $\mathcal{H}_{aux}^{\oplus}(\lambda)$ are not uniquely determined, different choices will give rise to unitarily equivalent Hilbert spaces. Crucially, we can show that, for such a direct integral decomposition, we will have that $\hat{\mathbf{M}}$ acts on $\mathcal{H}_{aux}^{\oplus}(\lambda)$ by multiplication by $\lambda$. We can then define a physical Hilbert space $\mathcal{H}_{phys}:=\mathcal{H}_{aux}^{\oplus}(0)$ which automatically comes equipped with a well defined inner product and upon which we can consider, if the uniform limit exists, a prospective class of observables in terms of the \textit{ergodic mean} of the $\hat{f}$, :
\begin{equation}
[\hat{f}]=\lim_{T\to \infty}\frac{1}{2T}\int_{-T}^{T}dte^{it\hat{\mathbf{M}}}\hat{f}e^{-it\hat{\mathbf{M}}}
\end{equation}
The existence of this object is guaranteed by the Birkhoff ergodic theorem since the unitary evolution operator, $U(t)=e^{it\hat{\mathbf{M}}}$, is a one-parameter measure preserving transformation on the Hilbert space (see \cite{Walt1981} \S1.6). From this definition we have that the $[\hat{f}]$ will both persevere the physical Hilbert space and induce a self-adjoint operator on that space. Furthermore, provided the spectral projections of the bounded operator $[\hat{f}]$ commute with those of $\hat{\mathbf{M}}$ (which we may expect) the ergodic mean can be seen to constitute a member of the class of \textit{strong observables}, $\hat{\mathcal{O}}_{s}$. These observables are defined to be functions such that relevant commutator with the master constraint vanishes identically: $[\hat{\mathcal{O}}_{s},\hat{\mathbf{M}}]\equiv 0$. Strong observables form a sub-set of the \textit{weak} observables $\hat{\mathcal{O}}$ defined via the quantum equivalent of the condition given by equation (3). 

It is important to note that despite the impressive improvements in formalising the Dirac quantisation programme that the RAQ and MCP approaches enable, these advances have come at the cost of removing our quantum formalism far from physical intuition. The most obvious way to ensure we have constructed more than just a mathematical edifice would be to demonstrate that both schemes have the appropriate classical limit. In particular, a proof that the  quantum observables reproduce their classical analogues in the appropriate limit would be highly desirable. Alternatively, one might seek to anchor these `quantise first, reduce second' techniques by a formal, or at least conceptual, correspondence with the less intuitively opaque `reduce first, quantise second' alternatives. It is to this task that we now turn.        

\subsection{Does quantisation commute with reduction?}

Provided we assume that the Poisson bracket algebra of the classical constraints closes with structure constants (and several further restrictions are imposed), we can often prove formally that a Dirac type procedure of quantising and then imposing the constraints at the quantum level is equivalent to first symplectically reducing and then geometrically quantising\footnote{This is a specific case of what is commonly referred to as the Guillemin-Sternberg conjecture \cite{GS1982}. See (for example) \cite{Gotay1986,BGMW1995,ConFre2009} for proofs of various degrees of generality}. The crucial results established  in such \textit{commutation} proofs is that: i) the physical Hilbert space constructed through a Dirac type approach, $\mathcal{H}_{phys}$, can be shown to be unitarily isomorphic to that (i.e. $\mathcal{H}_R$) achieved by quantising the symplectic manifold constructed by a classical reduction of the action of the constraints; and ii) the two quantization procedures result in an equivalent set of observables to the extent that the isomorphism in i) also intwines the representations of the two sets of quantum observables (both of which can be connected back to the same set of gauge invariant classical observables). We can thus assert, in certain circumstances, that quantisation does commute with reduction and assert physical equivalence in a strict sense.

Unfortunately the constraints of canonical general relativity are of such formidable complexity that the theory lies well outside any of the existent commutation proofs. Our purpose here will not be to examine exactly why this is the case nor to present work towards a solution of this very important problem. Rather, we will introduce a conceptual notion of commutativity which can be established in cases where formal arguments are not available. Such a weaker notion of commutativity will allow us to tackle the important task of exploring the somewhat unclear conceptual foundations of quantum theories constructed via a Dirac type methodology. In particular, it will give us a basis in which to examine the extent to which mathematical structures which they present us with are in correspondence with a reasonable physical ontology.  

To establish a conceptual notion of commutativity we do not need to look for the existence of a suitable observable intwining isomorphism existing between $\mathcal{H}_{phys}$ and $\mathcal{H}_R$ but rather establish  correspondence between both the Hilbert spaces and the observables on a representative level. We can flesh this idea out in terms of how the two approaches treat the relevant symmetries, observables and degrees of freedom of a given theory. What interests us is the extent to which the imposition of the constraints at a quantum level should be understood as implementing the same reduction from an otiose to a unique representative structure that we enact via classical reduction. The key features of the classical symplectic reduction procedure that we must require to be replicated at a quantum level are: 1) quotienting by the same gauge group; 2) reduction by the same number of degrees of freedom; and 3) the quantum observables defined via the two routes are equivalent to the extent that we are justified in thinking of them as representing the same underlying ontology. If we are satisfied as to equivalence in these three senses then we are justified in asserting that the two quantisation procedures produce \textit{representatively equivalent} structures and that representative commutation between quantisation and reduction holds. Clearly, this notion of commutativity in terms of representative equivalence should be implied by that defined in terms of unitary isomorphism but not visa versa. However, its potential significance is not \textit{purely interpretive} since, given a case where we believe the classical reduction to lead to physically unrealistic reduced phase space, establishing representative commutation will then provide us with grounds to doubt the physical basis of the theory quantised along the Dirac lines. Thus, what we have constructed is a heuristic tool as well as an interpretative criterion of equivalence.

Let us consider the case of a theory where: i) the constraints are associated with a Poisson bracket algebra with structure constants such that we can represent their action quantum mechanically via a set of unitary operators on an auxiliary Hilbert space; and ii) classically we can construct a reduced phase space with a symplectic geometry and a non-trivial Hamiltonian operator. In these circumstances, we can apply RAQ to produce a quantum theory via the Dirac type approach or alternatively proceed with a geometric quantisation of the reduced space. The key to evaluating our notion of representative commutativity is to examine the degree of correspondence between the quantum and classical reduction procedures. Immediately, we can see a \textit{prima facie} correspondence between the classical gauge orbits defined by the constraints on the physical phase space and the \textit{quantum orbits} defined by the U(g) on the auxiliary Hilbert space.  Further to this, we can also see an intuitive correspondence between a) the quotienting of the orbits in the classical theory to enable passage to the reduced space and b) the group averaging over the \textit{quantum orbits} that is used to construct the rigging map which projects into the physical Hilbert space. However as pointed out by Corichi \cite{Cor2008} we must be wary of taking these resemblances too seriously. Unlike in the classical case, the orbits are not generically equivalence classes of physical states -- this is to be expected since in the quantum case we do not make any restriction to a physical yet degenerate sub-space of the auxiliary Hilbert space which would be analogous to the physical phase space. Furthermore, the rigging map defined by group averaging will -- unlike the map to the reduced phase space -- in general take us to a state which is not part of the \textit{quantum orbit}. Thus, the two quotienting procedures are clearly different in an important sense. 

Nevertheless, despite these differences the two procedures \textit{are} equivalent in terms of quotienting out of the same gauge group. Since the constraints form a Lie algebra they are associated with a Lie group, $G$, at a classical level. It is the action of this group that we are removing from the physical phase space via symplectic reduction. This is the same group that we represent in RAQ in terms of unitary operators on $\mathcal{H}_{aux}$ and that we quotient out via the rigging map defined by group averaging in order to construct $\mathcal{H}_{phys}$. 

To see the correspondence in terms of degrees of freedom reduction we have to consider the nature of the group averaging procedure a little more carefully. Clearly, if all we were doing in RAQ was the quantum equivalent of classical reduction on an unconstrained space then we would have a mismatch in terms of number of degrees of freedom removed -- in the classical procedure half the excess degrees are removed by restriction to the physical phase space and half by the reduction itself. Rather, we must be able to understand the group averaging procedure as achieving both steps at once. We have, in fact, already considered the essence of the answer -- the rigging map does not just reduce out equivalence classes it projects onto physical states. If we start out with an unphysical state then it will take us to a physical state. If we start out with a solution to the constraints then, because the orbit is trivial, group averaging will keep us at the same point. Thus, as well as quotienting out of the same gauge group we also have the quantum equivalent of restriction to the physical phase space and the desired correspondence in terms of degrees of freedom reduction is guaranteed.   

Since the quantum observables of the RAQ scheme are defined such that they commute with the group action on the relevant sub-space of $\mathcal{H}_{aux}$, there is also a clear intuitive relationship between them and the classical observables on the physical phase space $\mathcal{O}$: both are in a sense `constant along the gauge orbits' -- although of course as we have seen the quantum gauge orbits are of a very different character to the classical ones. A stronger relationship  can be established between the observables on the reduced phase space $\mathcal{O}_{R}$ and the quantum observables of RAQ since both are well defined with respect to a non-degenerate and physical representative structure (i.e. $\Pi_{R}$ and $\mathcal{H}_{phys}$). In this context, we can then consider the associated quantum observables $\mathcal{\hat{O}}_{R}$ defined on $\mathcal{{H}_{R}}$ and (from an intuitive perspective) assert equivalence between them and the observables of RAQ such that we would be justified in thinking of them as representing the same underlying ontology. It should be noted that crucial to the establishment of this equivalence is the correspondence between the way \textit{states} are represented in $\Pi_{R}$ and $\mathcal{H}_{phys}$ respectively: it is because points in the respective spaces can be given analogous representative roles that we can establish a relationship between the observables defined via functions/operators on the spaces. 

It would therefore seem clear for the class of theories within which RAQ and symplectic reduction are applicable,  representative commutativity of reduction and quantisation will hold. The important question of whether our condition also holds for theories within which RAQ is not applicable, and the master constraint programme for quantisation has been applied, will be considered in the context of the Hamiltonian constraints of canonical general relativity in \S3.4.  

\section{Canonical general relativity; reduction and quantization}
\subsection{The ADM formalism and the constraint algebra}

For the class of globally hyperbolic solutions we can re-cast the original Lagrangian formulation of general relativity in vacuo into a constrained Hamiltonian formalism \cite{ADM1962}:
    \begin{eqnarray}
S=\frac{1}{\kappa}\int_{\mathbb{R}}dt\int_{\sigma}d^{3}x\{\dot{q}_{ab}P^{ab}-[N^{a}H_{a}+|N|H]\}
\end{eqnarray}

where $\kappa$ is the gravitational coupling constant, $\sigma$ is a three dimensional manifold of arbitrary topology, $q_{ab}$ and $P^{ab}$ are tensor fields defined on $\sigma$ and $N$ and $N^{a}$ are arbitrary multipliers called the lapse and shift. $H_{a}$ and $H$ are constraint functions of the form: 
    \begin{eqnarray}
H_{a}&:=&-2q_{ac}D_{b}P^{bc}\\
H&:=&-\frac{s\kappa}{\sqrt{det(q)}}[q_{ac}q_{bd}-\frac{1}{2}q_{ab}q_{cd}]P^{ab}P^{cd}-\sqrt{det(q)}\frac{R}{\kappa}
\end{eqnarray}
with $R$ the 3-scalar curvature and $s$ is the signature of the four metric in the Lagrangian theory we started off with (i.e. $-1$ for Lorentzian and $+1$ for Euclidean). The constraints are called the momentum and Hamiltonian constraints respectively. The physical phase space $\Sigma$ is a sub-manifold within the extended phase space $\Gamma$ defined by the constraints: 
\begin{eqnarray}
\Sigma =\{(q_{ab},P^{ab})=x\in \Gamma | H_{a}(x)=0;H(x)=0\}
\end{eqnarray}
Like in a typical gauge theory it has a presymplectic geometry $(\Sigma,\omega)$. 

The need for the imposition of these constraints is easily understood on an intuitive basis since we know that the physical modes of the classical gravitational field should correspond to a canonical representation with $4\times\infty^3$ degrees of freedom and these constraints serve to cut the $12\times\infty^3$ variables of the extended phase space down to $8\times\infty^3$ 
on the physical phase space. This still leaves another $4\times\infty^3$ unphysical degrees of freedom which we would normally seek to eliminate (classically) through symplectic reduction. Alternatively, we could seek to put off reducing out {\em all} of the excess degrees of freedom until after quantisation -- this would be to follow the Dirac route of \S2.2. The key issue is whether enacting the classical reduction for the case of the Hamiltonian constraints of canonical general relativity makes sense, and then whether the corresponding quantum reduction should be understood as having the same consequences. But before we bring our focus upon these questions in the context of the momentum and Hamiltonian constraints in turn it is crucial for us first to consider the structure of the constraint algebra.   

This algebra is know as the Dirac-Bergman algebra, $\mathfrak{D}$, (see \cite{Thie2007} \S 1.4 and references therein) and has an interesting and important structure. Explicit calculation of the relevant Poisson brackets gives: 
\begin{eqnarray}
\{\vec{H}(\vec{N}),\vec{H}(\vec{N'})\}&=&-\kappa \vec{H}(\mathfrak{L}_{N_{a}}N'_{a})\\
\{\vec{H}(\vec{N}),H(N)\}&=&-\kappa H(\mathfrak{L}_{N_{a}}N)\\
\{H(N),H(N')\}&=&s\kappa \vec{H}(F(N,N',q))
\end{eqnarray}
where $H(N)$ and $\vec{H}(\vec{N})$ are smeared versions of the constraints (e.g. $\vec{H}(\vec{N}) :=\int_{\sigma}d^{3}xN^{a}H_{a}$) and $F(N,N',q)=q^{ab}(NN'_{,b}-N'N_{,b})$.  Since it is phase space dependant $F(N,N',q)$ is a structure function and its presence on the right hand side of (9) indicates that the transformations associated with the algebra do not form a Lie group. Since the symmetries of the Lagrangian theory \textit{do} constitute a Lie group (that of four dimensional diffeomorphisms) we have an immediate worry that the canonical formalism may not have implemented some crucial structure of general relativity. However, under the condition that the equations of motion hold a correspondence can be established between the Lagrangian and Hamiltonian symmetries meaning that physically the two formulations of general relativity are indistinguishable. For our purposes this \textit{on shell} restriction as to the association of the constraints with diffeomorphism symmetries will prove crucial - it indicates that, unlike in a `typical' gauge theory (e.g. electromagnetism or Yang-Mills) the canonical generator of local symmetry transformations should be thought of as acting on \textit{entire solutions} and not on individual phase space points \cite{PonSalSun2010}. Proper appreciation of this non-standard facet of canonical general relativity is key to correctly interpreting the classical Hamiltonian constraints, and this classical interpretational question is in turn key to ensuring our quantisation procedures are conceptually coherent.     

\subsection{Reduction and quantisation of the momentum constraints}

Let us now focus on the momentum constraints in particular. The equation (7) above closes with structure constants and this means that the action of the momentum constraints can be associated with a Lie group. Moreover, this group can be understood explicitly in terms of the implementation of a Lie algebra of diffeomorphisms of the space-like hypersurface $\sigma$ \cite{IshKuc1985}. This solid geometrical interpretation of the action of the constraints is further illustrated by considering the explicit form of the Poisson brackets between the constraints and arbitrary canonical variables on the extended phase space.  

\begin{eqnarray}
\{\vec{H}(\vec{N}),q_{ab}\}=\kappa(\mathfrak{L}_{\vec{N}}q_{ab})\\
\{\vec{H}(\vec{N}),P^{ab}\}=\kappa(\mathfrak{L}_{\vec{N}}P^{ab})
\end{eqnarray}  
The appearance of the Lie derivative on the right hand side of each equation indicates that these constraints can be understood as purely generating infinitesimal diffeomorphisms. We would then seem fully justified classically in seeking to: i) quotient the action of these constraints via the application of symplectic reduction; and ii) construct a partially reduced phase space where each point will correspond to canonical variables defined upon a spatially diffeomorphic invariant three geometry. Such a space is the cotangent bundle associated with Wheeler's superspace \cite{Wheeler1968} and as such we shall call it the super-phase space, $T^{\star}\mathcal{S}$. Formally, its structure is little explored and it is unlikely to be without singularities and other topological complications.\footnote{See \cite{Giu2009} for detailed discussion of the metric and topological structure of $\mathcal{S}$} However, from a conceptual viewpoint its representational role is clear and we will therefore make the (highly non-trivial) assumption that it has the characteristics of a typical reduced phase space with the associated symplectic geometry. As such, the application of geometric quantisation would be available and a corresponding Hilbert space $\mathcal{H}_{SPS}$ could be constructed. 

For our purposes what is most significant is what representational relationship such a Hilbert space would have to that constructed via a Dirac type `quantise first, reduce second' route. Since the momentum constraints are associated with an algebra which closes with structure constants, it would seem appropriate to think of the associated Lie group as being representable quantum mechanically in terms of the action of a set of unitary operators on an auxiliary Hilbert space. Unfortunately, there is complication here since within modern approaches (i.e. LQG) it is found that we are in fact only able to construct quantum operators generating the finite component of the spatial diffeomorphism group. Although some variant of the group averaging methodology of the RAQ scheme can then be applied\footnote{See \cite{Thie2007} \S9 for extensive details of such a methodology for the Dirac type quantisation of the momentum constraints in the context of Ashtekar variables}, this will lead us ultimately to to a non-separable physical Hilbert space $\mathcal{H}_{mom}$. 

The relationship between $\mathcal{H}_{mom}$ and $\mathcal{H}_{SPS}$ (which we would assume to be separable) is not going to be simple. Formally, the two spaces are certainly not going to be unitarily isomorphic and even representationally we do not have an exact correspondence since the groups involved in the classical and quantum quotienting procedures are strictly speaking different. However, in terms of the putative ontology represented by these two Hilbert spaces, these details are not crucial. The classical and quantum quotients are equivalent in that both lead us to the representation of objects invariant under spatial diffeomorphisms (albeit in slightly different sense since in the former but not in the latter case the diffeomorphisms are smooth). Furthermore, in terms of degrees of freedom we will have equivalence since in both cases we are cutting down by $6\times\infty^3$. On the level of observables too we can argue towards equivalence since a representation of an algebra of spatially diffeomorphism invariant observables is well defined on $\mathcal{H}_{mom}$. Clearly, it is reasonable  to think of such an algebra as representing the same fundamental objects as the $\mathcal{\hat{O}}_{R}$ which we would define on $\mathcal{{H}_{SPS}}$. 

Thus, although we can not strictly assert representational commutation between reduction and quantisation for the momentum constraints -- because of the problems in constructing a quantum operator which generates infinitesimal spatial diffeomorphisms -- we can assert commutation to hold \textit{for all intents and purposes} since one may at least represent the same spatially diffeomorphism invariant ontology via both Dirac and reduced quantisation routes.

\subsection{Classical Hamiltonian constraints, reduction and the problem of time} 

As mentioned above the Poisson bracket between Hamiltonian constraints closes only with structure functions. This means their interpretation resists a simple group theoretic basis. Connectedly we also have that the transformations generated by all the constraints together can only be understood as implementing a canonical representation of the fundamental local symmetry group of the Lagrangian theory when a solution to the equation of motion has already been defined. This peculiar feature of the theory is directly related to the structure of the Hamiltonian constraint as is illustrated by the investigation of its action on an \textit{embedded} canonical momenta variable. Such a variable is so called because it is the canonical conjugate of an metric variable $q^{\mu\nu}$ which is a tensor field (the first fundamental form) defined on the embedded hypersurface $\Sigma_{t}$. This new metric variable can be expressed purely in terms of spatial vector fields on $\Sigma_{t}$ and the usual metric variable on $\sigma$, $q^{ab}$ (see \cite{Thie2007} (1.1.16)) and the new momenta variable can be written in terms of it together with it and another spatial tensor field on $\Sigma_{t}$ (the second fundamental form). An elegant calculation by Thiemann (\cite{Thie2007} pp.54-56) yields the explicit expression :
\begin{eqnarray}
\{H(N) , P^{\mu\nu}\} &=& \frac{q^{\mu\nu} NH }{2} - N \sqrt{\|q\|} [q^{\mu\rho} q^{\nu\sigma} - q^{\mu\nu}q^{\rho\sigma}] R_{\rho\sigma}^{4} + \mathfrak{L}_{Nn}P^{\mu\nu}
\end{eqnarray}
with $R_{\mu\nu}^{4}$ the Ricci 4-tensor. The first term on the right hand side is zero on the physical phase space and is therefore unimportant. The second is zero for solutions to the equations of motion and thus by the third meaning term we have that the Hamiltonian generates infinitesimal diffeomorphisms \textit{on shell}. However, this interpretation only holds in the context of spatial hypersurface embedded in a dynamical solution -- it does not give us an unambiguous understanding of  the action of $H(N)$ on the three dimensional canonical variables that parameterise the physical phase space. In particular, the transformations between the three geometries $\sigma$ which collectively constitute the solutions are themselves also generated by the phase space action of the Hamiltonian constraint. This implies that the Hamiltonian constraints of canonical gravity are of an unusual \textit{dynamical} type such that the transformations they effect on the physical phase space cannot be understood purely as gauge. Furthermore, it also indicates that sticking with such a \textit{pure gauge} interpretation will involve forgoing the constraints role in the generation of dynamics.\footnote{This highly significant, and yet ill-appreciated, point is at the heart of the Kucha\v{r}-Barbour critique \cite{Kuch1992,Bar1994} of the standard \textit{Dirac dogma}  (that all first class constraints are gauge generating) as applied to the Hamiltonian constraints of canonical general relativity.}

This should be of no surprise since similar constraints which arise in the canonical formulation of both parameterised particle dynamics and Jacobi's theory can also be shown to have such a dynamical character \cite{BarFos2008}. As I have argued elsewhere for those simpler cases \cite{The2010}, the presence of constraints with such a dynamical aspect implies that the application of symplectic reduction will have a trivialising effect such that any reduced phase space will not have the capacity for the representation of dynamics. We can see this explicitly for the case of general relativity in particular: I) As mentioned above the physical phase space of the theory will have a presymplectic geometry $(\Sigma,\omega)$ and like in a \textit{typical} gauge theory this structure contains characteristic `null directions' the `integral curves' of which we usually identify as gauge orbits. II) Since general relativity is a field theory the geometric structure we are dealing with is a little more complex than that introduce in \S2.1: motions are now four-dimensional surfaces with a quadritangent $X$ (made up of the tensor product of four independent tangents) defined at each point. Thus, the part of `null direction' is now played by the a specific quadritangent and that of `integral curve' by an integral surface. III) More significantly, and in stark contrast to the typical case, if we define the gauge orbits of canonical general relativity explicitly in terms of the four dimensional surfaces $\bar{\gamma}$ in $\Sigma $ such that the quadritangent to the orbit $X$ is in the kernel of $\omega $ (i.e. $\omega (X)=0$) then we can identify the $\bar{\gamma}$ with the set of (globally hyperbolic) solutions of the Einstein field equations \cite{Rov2004}. IV) Since these orbits are precisely those which we would normally classify as gauge equivalence classes a symplectic reduction procedure would (in principle) lead to a reduced phase space within which, \textit{prima facie}, dynamics has been gauged out. V) Furthermore, since this reduced space is only equipped with a trivial Hamiltonian function there is no hope of recovering dynamical evolution in terms of transformations between points in the reduced space.   

In light of these arguments, one might then attempt to re-interpret the reduced space as a space of histories with each point taken to represent a solution invariant under the class of four-dimensional diffeomorphisms \cite{Belot2007}. This could be justified on the basis that there exists a single canonical isomorphism from our reduced phase space points to the space of gauge invariant solutions in a Lagrangian formalism. However, the existence of an isomorphism does not automatically confer representational equivalence and if we read the reduced space in such a manner then it may have problematic consequences for how we view the unreduced phase space. In fact, since such an interpretation would lead us towards treating the gauge orbits on the physical phase space as equivalence classes of diffeomorphically related four geometries it is inconsistent with the very basis of the ADM decomposition of space-time into three geometries. Rather, since the representational role of points in the ADM phase space is clearly predicated upon a three dimensional ontology, points within the symplectic reduction of this space are required to perform an analogous role. Thus, the absence of a non-zero Hamiltonian function means that reduced phase space is inherently trivial as a structure for representing dynamics -- the only ontology it can be associated with is that of static three geometries.  

The vital realisation is that we can only view the constraints of canonical general relativity as collectively producing four dimensional diffeomorphisms once a dynamical solution generated by the Hamiltonian constraint has already be defined. Thus the Hamiltonian constraint's action on the phase space is essentially dual and passing to the reduced phase space of canonical gravity (where the action of the Hamiltonian constraint is treated as pure gauge) will inevitably involve throwing the dynamical baby out with the diffeomorphism symmetry bathwater. The important lesson that we should learn is not that considering general relativity in terms of a canonical formalism leads inevitably to a `frozen' picture of the world represented in terms of a dynamically trivial reduced space -- this way of portraying the problem of time is misguided. Rather, the fundamental point is that blind application of the standard interpretation of first class constraints, and the symplectic reduction that goes along with it, to the Hamiltonian constraints leads to a physically unreasonable formalism.


\subsection{How should we interpret the quantum Hamiltonian constraints?}

The preceding discussion leads us directly to the essential dilemma -- should we understand the implementation of the Hamiltonian constraints in terms of operators annihilating the wavefunction according to the Dirac quantisation prescription as equivalent to the conceptually problematic classical reduction? Or more precisely, is it appropriate to think of reduction and quantisation as commutative procedures when considered with regard to the Hamiltonian constraints? On a formal level, it is not yet possible to answer this question since the Hamiltonian constraints lie outside the scope of existent commutation proofs. Furthermore, we cannot at the moment even make use of our weaker representative notion of commutativity since we have only established its viability for cases in which the constraints close with structure constants and the RAQ refinement of Dirac quantisation is available. We can at least argue towards some degree of representative equivalence between the naive quantisation of the Hamiltonian constraint via the original Dirac quantisation methodology (leading to the Wheeler-De Witt equation) and a quantisation of the putative and problematic reduced phase space since there is an equivalence in terms of reduction of degrees of freedom by $2\times\infty^3$. However, since the original Dirac does not guarantee us either a well defined physical Hilbert space nor a set of observables and there is no group theoretic basis for interpreting the relevant symmetries, we are still well short of securing even our weak notion of representative equivalence.       

Rather, quantisation of the Hamiltonian constraints\footnote{We should here, more properly, be speaking of the Hamiltonian constraints as reformulated in Ashtekar variables rather than those expressed in normal ADM variables. However, since the reformulated Hamiltonian constraints close with the same Poisson bracket structure (as they must), this difference is immaterial to our current purpose -- although it will become important within a more explicit treatment.} is the context within which the master constraint programme comes into its own. Dittrich and Thiemann \cite{Thie2007,DitThe2006}  have produced encouraging results with regard to the applicability of this scheme to the Hamiltonian constraints (although the significant problem establishing the correct classical limit, among others, still remains) and it would therefore seem, in the first instance, reasonable to assume that if we can establish in general the viability of representational commutation for theories in which MCP has been applied, then we have a good basis for representational commutation in the case of the Hamiltonian constraints.   

Recall from above that in the MCP we seek classically to construct a single \textit{master constraint} the satisfaction of which is equivalent to the satisfaction of all the individual constraints. We then promote this single constraint to a self-adjoint operator on an auxiliary Hilbert space and then use the direct integral methodology to construct a well defined physical Hilbert space. To establish representational commutativity we first look to find a correspondence between the classical and quantum reductions in terms of reduction by the same number of degrees of freedom. We can do this be considering the quantum master constraint equation $\hat{\mathbf{M}}\psi=0$ which we implicitly solve when constructing the physical Hilbert space via the direct integral method. Following Thiemann \cite{Thie2007} we can consider the simple case that $\hat{\mathbf{M}}=K_{i}\hat{\varphi _{i}}^{\dag}\hat{\varphi _{i}}$ where $K_{i}>0$ are constants with the required convergence properties.\footnote{Here the $\hat{\varphi _{i}}$ are a countable and close-able set of operators which need not be self-adjoint nor form a Lie algebra but are such that $\{0\}$ lies only in their common point spectrum.} Next we have that $\hat{\mathbf{M}}\psi=0$ implies that $\hat{\varphi} _{i}\psi=0$ since by definition $\langle\psi\mid\hat{\mathbf{M}}\psi\rangle=K_{i}\|\hat{\varphi} _{i}\psi\|^{2}=0$. We can then fall back on the correspondence \cite{HenTeit1992} in terms reduction of degrees of freedom between the Dirac quantum constraint conditions $\hat{\varphi} _{i}\psi=0$ and the classical symplectic reduction of a system with physical phase space $\Sigma =\{(p,q)\in \Gamma|\forall _{i}:\varphi _{i}(p,q)=0\}$.        

Moving on to the condition regarding observables: we have from above that the MCP allows us to define the strong observables $\hat{\mathcal{O}}_{s}$ which are such that $[\hat{\mathcal{O}}_{s}, \hat{\mathbf{M}}]\equiv 0$. What kind of relationship is there between such observables and the $\hat{\mathcal{O}}_{R}$ that we construct based upon the reduced phase space? We can address this question by first considering the \textit{weak} classical observables which were defined by the double commutator $\{\{\mathbf{M},\mathcal{O}\},\mathcal{O}\}|_{\mathbf{M}=0}=0$. We have that $\{\{\mathbf{M},\mathcal{O}\},\mathcal{O}\}|_{\mathbf{M}=0}=0$ is equivalent to $\{\varphi _{i},\mathcal{O}\}|_{\mathbf{M}=0}=0$. This means that, as noted above, we can think of a geometrical correspondence between $\mathcal{O}$ and $\mathcal{O}_{R}$ since the first are constant along the gauge orbits which are quotiented out in order to construct the space in which the latter are defined. Since the classical strong observables (which can be constructed by considering an ergodic mean analogous to (5)) are a sub-set of the weak observables such a correspondence will hold for them also, and it seems correct to think of $\hat{\mathcal{O}}_{s}$ as being representatively equivalent to a sub-set of the $\hat{\mathcal{O}}_{R}$.  Classically, we can in fact give a formal criterion to define this sub-set since they will be such that the pull-back of the map which projects down to the reduced space (i.e. $\pi^{\star} :\Pi _{R}\rightarrow\Sigma$) will take them to the $\mathcal{O}_{s}$. We have not yet considered the representation of the physical state space over which these observables are defined and, as was pointed out above, this relationship is in fact ket to establishing representative correspondence between the observables. We will return to this issue at the end of this section.     

More problematic is our condition concerning `quotienting by the same gauge group' -- since MCP is still well defined for cases (such as that of the Hamiltonian constraints) where there is no group theoretic basis to the quotient taken in symplectic reduction, the condition clearly must be adapted to remain relevant. Instead, we should look for the same set of local transformations being removed without any restriction on the nature of these transformations (i.e they may not form a group). Let us return our focus to the Hamiltonian constraints of canonical general relativity. The crucial question is then whether we should understand the MCP as enacting a quantum equivalent of the dynamically trivialising classical reduction discussed in \S3.3. In particular, are we doing something equivalent to erroneously treating the (at least) partially dynamical action of the constraints purely as a gauge transformation on the physical phase space? Let us consider the explicit form of the master constraint for canonical general relativity \cite{Thie2007,Thie2003,DitThe2006}: 
\begin{eqnarray}
\mathbf{M}=\frac{1}{2}\int_{\sigma}d^{3}x\frac{H(x)^{2}}{\sqrt{det(q)}(x)}
\end{eqnarray}
This constraint has a number of formal virtues. In particular it is such that its satisfaction implies that $H(N)=0$ for all $N$ meaning that encodes the same constraint surface as the Hamiltonian constraints. Furthermore, it is also such that $\{\vec{H}(\vec{N}),\mathbf{M}\}=0$ meaning that it is invariant under spatial diffeomorphisms and will lead us to a constraint algebra with a much simpler form, the \textit{master constraint algebra} $\mathfrak{M}$:
\begin{eqnarray}
\{H_{a}(N_{a}),H_{a}(N'_{a})\}&=&-\kappa H_{a}(\mathfrak{L}_{N_{a}}N'_{a})\\
\{H_{a}(N_{a}),\mathbf{M}\}&=&0\\
\{\mathbf{M},\mathbf{M}\}&=&0
\end{eqnarray}
We no longer have to deal with the presence of structure functions in our constraint algebra since the highly complex expression (12) in the Dirac algebra is replaced by the trivial self-commutation expression (19). In substituting a single master constraint for the infinite set of Hamiltonian constraints we avoid having to explicitly confront the difficulties of the Poisson bracket algebra with which the latter are associated. Furthermore, since the master constraint algebra \textit{is} a proper Lie algebra it can of course be associated with a Lie group of transformations. This means that the task of fully quantising canonical general relativity (i.e. dealing with both sets of constraints) will be made far more tractable. Returning to the point in hand, clearly $\mathfrak{M}\neq\mathfrak{D}$ since one is a Lie algebra and one is not. So there is a clear sense in which symplectic reduction (which removes the action of the transformations associated with $\mathfrak{D}$) is not going to have a straight forward representational relationship to application of the MCP. Yet, we were able to establish a degree of correspondence in terms of the treatment observables so we should still expect \textit{some} correspondence in terms of which transformations the two reductions treat as unphysical. We might hope to get a definite formal grip on this relationship by calculating the action of $\mathbf{M}$ on a phase space variable. However, since such a calculation will only yield an expression which is vanishing for $\mathbf{M}=0$ it is clear that the action constructed in this way will be trivial on the physical phase space. The key realisation is that since the Hamilton vector field associated with the master constraint, $X_{\mathbf{M}}^{a}$, is by definition vanishing on the physical phase space the Poisson bracket between it and any phase space function will always be zero for $\mathbf{M}=0$. Thus, there are no interpretational difficulties in treating the orbit associated with the integral curves of $X_{\mathbf{M}}^{a}$ as gauge  since it is a trivial move.

To make more definite progress we must consider the quantum theory. Recall from above that we look to represent the master constraint as a positive, self-adjoint operator $\hat{\mathbf{M}}$ on an auxiliary Hilbert space $\mathcal{H}_{aux}$. We then use the direct integral methodology to construct a physical Hilbert space, $\mathcal{H}_{phys}$. Setting aside some important technical complications not least the non-separability of $\mathcal{H}_{aux}$ (see \cite{Thie2007} \S10.6.3 on this point), the essential elements of this scheme are readily applicable to our master constraint formulation of the Hamiltonian constraints of classical general relativity. What is important for our purpose is whether in constructing $\mathcal{H}_{phys}$ we have carried out a move analogous to treating the classical action of the Hamiltonian constraints on phase space as pure gauge. At first sight, it appears that we have not since the \textit{quantum quotient} that we take in order to construct $\mathcal{H}_{phys}$ is with respect to the kernel of $\hat{\mathbf{M}}$. 

Considering things more carefully, the direct integral methodology represents $\hat{\mathbf{M}}$ on $\mathcal{H}_{aux}^{\oplus}(\lambda)$ such that 
\begin{equation}
\hat{\mathbf{M}}(\psi_{aux}(\lambda))_{\lambda\in\mathbb{R}}=(\lambda\psi(\lambda))_{\lambda\in\mathbb{R}}
\end{equation}
and then \textit{defines} $\mathcal{H}_{phys}$ in terms of the $\psi_{aux}(\lambda)$ in $\mathcal{H}_{aux}^{\oplus}(\lambda)$ which are such that $\lambda$ equals zero. This of course means that only states which solve the master constraint will be part of the physical Hilbert space. Furthermore it also means that (following Corichi \cite{Cor2008}) we should think of the quantum equivalent to the Hamilton vector field of the master constraint as vanishing.\footnote{Our ability to apply these classical geometrical terms in the quantum context  derives from the symplectic structure encoded in the space of rays associated with any Hilbert space. See \cite{Cor2008} and references therein for more details} In fact, since the master constraint can be represented in terms of a positive self adjoint operator on $\mathcal{H}_{aux}$, $\hat{\mathbf{M}}$ is associated with a one parameter family of unitary operators, $\hat{U}(t)=e^{it\hat{\mathbf{M}}}$. It is therefore appropriate to think of the construction of $\mathcal{H}_{phys}$ in terms of the quotienting of a quantum gauge orbit associated with $\hat{U}(t)$ in the same sense as we discussed for the case of RAQ. This would seem to indicate that our intuition from the classical theory has proved correct -- quantisation according to the MCP should not, when applied to the Hamiltonian constraints, be consider as involving a \textit{quantum quotienting} analogous to that achieved be reducing out the constraints at a classical level. 

We have, however, neglected to consider the Dirac observables -- it is only in virtue of them that the master constraint can be said to encode the same classical structure as the individual constraints. In fact, according to Thiemann \cite{Thie2003}, the requirement that both the observables and the individual constraint operators be represented as self adjoint operators on $\mathcal{H}_{phys}$, can be shown (in solvable models) to fix the inner product such that the solution space must be reduced to the simultaneous one of all constraints. This implies that states in the auxiliary Hilbert space which fail to be solutions of the \textit{individual constraints} will be excluded in the passage to the physical Hilbert space. If this were to hold for the Hamiltonian constraints of canonical general relativity then we would have a restriction on physical states such that they: i) individually solve Wheeler-De Witt type equations of the form $\hat{H}\psi_{phys}=0$; and ii) collectively solve the master constraint equation $\hat{\mathbf{M}}\psi_{phys}=0$.\footnote{Whether this proves to be the case in practice can only be established by a full treatment, with the quantum Hamiltonian constraints reformulated in terms of loop variables and the Dirac observables explicitly constructed (presumably using the \textit{partial observables Ansatz} \cite{Thie2007,Rov2004, Dit2006, Dit2007}).} Under these circumstances, we can then argue that the representation of physical states arrived at via this `quantise first, reduce second' methodology will coincide with that based upon quantisation of the dynamically trivial classically reduced space. This is because if the physical Hilbert space is such that only states which are zero eigenvectors of the Hamiltonian constraint operators are permitted, then no two distinct \textit{classical} states which lie along the null direction which the classical constraint function defines can be represented at the quantum level. This means that fundamentally the same set of objects have been excluded from our ontology as in the case of a reduced and then quantised theory. We would therefore be justified in asserting that the quotienting criteria of representational equivalence will hold since we have recovered its fundamental aspect.

Furthermore, this conceptual connection between the physical states also ensures that there is full representative correspondence between the reduced and physical observables at a quantum level and therefore that our criteria concerning observables holds. Thus, for both the general case and the specific case of general relativity the physical Hilbert space constructed via the MCP is representationally equivalent to that based upon quantisation of a reduced phase space -- i.e. representational commutation between quantisation and reduction holds. This gives us strong conceptual grounds for doubting the validity of applying this quantisation procedure to the Hamiltonian constraints of general relativity on the grounds of the trivialisation argument of \S 3.3. Since classically it is incoherent to treat the Hamiltonian constraints as purely generating unphysical phase space transformations, any approach that is equivalent to the implementation of this interpretation at a quantum level will be similarly afflicted.    

\section{Conclusion}

The establishment of representative commutativity between reduction and quantisation for the case of the Hamiltonian constraints of general relativity provides us with a significant conceptual insight into the canonical quantisation of general relativity. Since reduction as applied to the classical action of the Hamiltonian constraints is manifestly trivialising, any approach to quantum gravity that arrives at a physical Hilbert space that is representatively equivalent to the quantisation of this reduced space rests on unsolid ground. Quantisation of the constraints according to the master constraint programme (which is at the moment the only viable methodology) seems in principle to be such that the representation of the world it provides will coincide with that based upon the problematic classical reduction. If accepted, these arguments necessitate a revelation of the foundations upon which current approaches to the canonical quantisation of gravity are defined. 
         
\section*{Acknowledgements}
Thanks to Sean Gryb, Eilwyn Lim and two anonymous referees for helpful comments on earlier drafts of this paper, and to Hans Westman, Dean Rickles and Julian Barbour for numerous constructive discussions relating to Hamiltonian constraints and the problem of time.


\bibliographystyle{mdpi}
\makeatletter
\renewcommand\@biblabel[1]{#1. }
\makeatother

\begin{thebibliography}{1}
\bibitem{Dirac1964}
Dirac, P. A. M. {\em Lecture on quantum mechanics}; Dover: Mineola, NY, USA, 1964.
\bibitem{HenTeit1992}
Henneaux, M.; Teitelboim, C. {\em Quantization of Gauge Systems}; Princeton University Press: Princeton, NJ,  1992.
\bibitem{Butt2007} 
Butterfield, J. On symplectic reduction in classical mechanics. In {\em
Handbook of Philosophy of Physics}; Butterfield, J., Earman, J., Eds.; Elsevier, Oxford, UK, 2007, pp.1-132.
\bibitem{GotNesHin1978}
Gotay, M.J.; Nester, J.M.; Hinds, G. Presymplectic manifolds and the Dirac-Bergman theory of constraints. {\em 
J. Math. Phys.}{\bf 1978}, {\em 19}, 2388-99.
\bibitem{Sour1997}
Souriau, J. M. {\em Structure of dynamical systems: a symplectic view of physics}; Cushman-de Vries, C.H. Trans.; Cushman, R.H.,Tuynman, G.M., Trans. Eds.; Birkhauser: Cambridge, MA, USA 2007.
\bibitem{EEML1999} 
Echeverria-Enriquez A.; Munoz-Lecanda M. Mathematical foundations of geometric quantization. \href{http://arxiv.org/abs/math-ph/9904008} {\em arxiv.org/abs/math-ph/9904008} {\bf1999} 
\bibitem{Lands2007} 
Landsman, N.P. Between classical and quantum. In {\em
Handbook of Philosophy of Physics}; Butterfield, J., Earman, J., Eds.; Elsevier, Oxford, UK, 2007, pp. 417-554.
\bibitem{GS1982}
Guillemin, V.; Sternberg, S. Geometric quantization and multiplicities of group representations. {\em Invent. Math.} {\bf 1982}, {\em 67}, pp. 515-538. 
\bibitem{Gotay1986}
Gotay, M. Constraints, reduction, and quantization. {\em J. Math. Phys.} {\bf 1986}, {\em 27}, pp. 2051-66.  
\bibitem{BGMW1995}
Duistermaat, H.; Guillemin, V.; Meinrenken, E.; Wu, S. Symplectic reduction and Riemann-Roch for circle actions, {\em Math. Res. Letters} {\bf 1995}, {\em 2}, pp. 259-66.  
\bibitem{ConFre2009}
Conrady, F. and Freidel, L. Quantum geometry from phase space reduction. {\em J. Math. Phys.}  {\bf 2009}, {\em 50}, pp.123510-39.
\bibitem{Kuchar1991}
Kucha\v{r}, K. The problem of time in canonical quantization of relativistic systems. In {\em Conceptual problems of quantum gravity}; Ashtekar, A., Stachel, J., Eds.; Springer: New York, NY, USA, {\bf 1991}, pp. 141-171.
\bibitem{Isham1992}
Isham, C. Canonical quantum gravity and the problem of time. \href{http://arxiv.org/abs/grqc/9210011}{\em arxiv.org/abs/grqc/9210011} {\bf1992}
\bibitem{Ander2010}
Anderson, E. The Problem of Time in Quantum Gravity. \href{http://arxiv.org/pdf/1009.2157} {\em arxiv.org/pdf/1009.2157} {\bf 2010}
\bibitem{PonSalSun2010}
Pons, J. M.; Salisbury, D.C.; Sundermeyer, K. A. Observables in classical canonical gravity: folklore demystified, {\em J. Phy.: Conf. Ser.} {\bf 2010}, {\em 222}, 012018. 
\bibitem{Kuch1992}
Kucha\v{r}, K. Time and interpretations of quantum gravity. In {\em Proceedings of the fourth Canadian conference on general relativity and relativistic astrophysics}; Kunstatter, G., Vincent, D., Williams, J. , Eds.; World Scientific: Singapore; 1992, pp. 211-314.
\bibitem{Bar1994}
Barbour, J. The timelessness of quantum gravity: I evidence from the classical theory. {\em Class. Quant. Grav.} {\bf 1994}, {\em 1}, pp. 2853-74. 
\bibitem{Thie2007}
Thiemann, T. {\em Modern canonical quantum general relativity}; Cambridge University Press: Cambridge, UK, {\bf 2007}
\bibitem{GiulMar1999a}
Giulini, D.; Marolf, D. On the generality of refined algebraic quantization. {\em Class. Quant. Grav.} {\bf 1999}, {\em 16}, pp. 2479-88. 
\bibitem{GiulMar1999b}
Giulini, D.; Marolf, D. A uniqueness theorem for constraint quantization. {\em Class. Quant. Grav.}  {\bf 1999} , {\em 16}, pp.2489-506.
\bibitem{Thie2003}
Thiemann, T. The Phoneix Project: Master Constraint Programme. \href{http://arxiv.org/abs/gr-qc/0305080}{\em{arxiv.org/abs/gr-qc/0305080}} {\bf 2003}
\bibitem{DitThe2006}
Dittrich, B and Thiemann, T. Testing the master constraint programme for loop quantum gravity: I. General framework. {\em  Class. Quant. Grav.} {\bf 2006}, {\em 23}, pp. 1025-65. 
\bibitem{Walt1981}
Walters, P. {\em An Introduction to Ergodic Theory};  Springer-Verlag: New York, USA, {\bf 1981} 
\bibitem{Cor2008}
Corichi, A. On the geometry of quantum constrained systems. \href{http://arxiv.org/abs/0801.1119}{\em arxiv.org/abs/0801.1119} {\bf 2008}
\bibitem{ADM1962}
Arnowitt, R.; Deser, S.; Misner, C. W. The Dynamics of General Relativity. In {\em Gravitation: an introduction to current research}; Witten. L, Ed.; Wiley: New York, NY, 1962; pp.227-39 
\bibitem{IshKuc1985}
Isham, C.; Kucha\v{r}, K. Representations of Spacetime Diffeomorphisms. I Canonical Parametrized Field Theories. {\em Annals of Physics} {\bf1985}, {\em164}, pp. 288-315.
\bibitem{Wheeler1968}
Wheeler, J.A. Superspace and the nature of quantum geometrodynamics. In {\em Lectures in Mathematics and Physics}; DeWitt, C.M., Wheeler, Eds.; Benjamin: New York, NY, USA, 1968; pp. 242-307
\bibitem{Giu2009}
Giulini, D. The Superspace of geometrodynamics. {\em Gen. Relativ. Gravit.} {\bf 2009}, {\em 41}, pp. 785-815 
\bibitem{BarFos2008}
Barbour, J.; Foster, B. Constraints and gauge transformations: Dirac's theorem is not always valid. \href{http://arxiv.org/pdf/0808.1223}{\em arxiv.org/pdf/0808.1223} {\bf 2008}
\bibitem{The2010}
Thebault, K. P. Y. Symplectic reduction and the problem of time in nonrelativistic mechanics. \href{http://philsci-archive.pitt.edu/8433/}{\em philsci-archive.pitt.edu/8433/} {\bf 2010}
\bibitem{Rov2004}
Rovelli, C. {\em Quantum gravity}. Cambridge University Press: Cambridge, UK, 2004.
\bibitem{Belot2007} 
Belot, G. The representation of time and change in mechanics. In {\em
Handbook of Philosophy of Physics}; Butterfield, J., Earman, J., Eds.; Elsevier, Oxford, UK, 2007, pp.133-228.
\bibitem{Dit2006}
Dittrich, B. Partial and complete observables for
canonical general relativity. \textit{Class. Quant. Grav.} {\bf 2006}, {\em 23} pp. 6155-85.
\bibitem{Dit2007}
Dittrich, B. Partial and complete observables for
Hamiltonian constrained systems. \textit{Gen. Rel. Grav.} {\bf 2007}, {\em 39}, pp. 1891-927.

\end{thebibliography}

\end{document}